\documentclass[12pt,preprint]{aastex}

\usepackage{epsfig}

\lefthead{Zhang, Flyer & Low}
\righthead{Magnetic-Field Confinement: Helicity}

\begin{document}

\title{Magnetic Field Confinement in the Corona:
The Role of Magnetic Helicity Accumulation}

\author{Mei Zhang\altaffilmark{1,2},
Natasha Flyer\altaffilmark{3}
\& Boon Chye Low\altaffilmark{2}}

\altaffiltext{1}{National Astronomical Observatory,
Chinese Academy of Sciences, Datun Road A20,
Chaoyang District, Beijing 100012, China}
\altaffiltext{2}{High Altitude Observatory, National Center 
for Atmospheric Research, PO Box 3000, Boulder, CO 80307, USA}
\altaffiltext{3}{Scientific Computing Division, National 
Center for Atmospheric Research, PO Box 3000, Boulder, 
CO 80307, USA}


\begin{abstract}
A loss of magnetic field confinement is believed to be the
cause of coronal mass ejections (CMEs), a major form of
solar activity in the corona. The mechanisms for magnetic
energy storage are crucial in understanding how a field
may possess enough free energy to overcome the Aly limit
and open up. Previously, we have pointed out that the
accumulation of magnetic helicity in the corona plays a
significant role in storing magnetic energy. In this
paper, we investigate another hydromagnetic consequence of
magnetic-helicity accumulation. We propose a conjecture
that there is an upper bound on the total magnetic helicity 
that a force-free field can contain. This is directly related 
to the hydromagnetic property that force-free fields in 
unbounded space have to be self-confining. Although a
mathematical proof of this conjecture for any field
configuration is formidable, its plausibility can be
demonstrated with the properties of several families of
power-law, axisymmetric force-free fields. We put forth
mathematical evidence, as well as numerical, indicating
that an upper bound on the magnetic helicity may exist for
such fields. Thus, the accumulation of magnetic helicity
in excess of this upper bound would initiate a non-equilibrium 
situation, resulting in a CME expulsion as a natural 
product of coronal evolution.
\end{abstract}

\keywords{MHD --- Sun: magnetic fields --- Sun: corona
--- Sun: coronal mass ejections (CMEs) }


\section{Introduction}

Coronal mass ejections (CMEs) are a major form of
solar activity in the corona. A CME takes $10^{15-16}$ g
of plasma from the low corona into the solar wind
to disturb the near-Earth space if the CME propagation
is directed toward us. Unlike other solar activities such as
flares and filament eruptions, CMEs were only first observed
when white light coronagraphs were put into space in the
1970s (MacQueen et al. 1974, Howard 1976). Since then,
this astrophysical process has caught much attention in the
space physics community because of the adverse
consequences of CMEs for orbiting satellites.

CMEs are identified as large-scale bright features
in white light that move outward through the corona 
at speeds from 10 to over 2000 km/s (Hundhausen 1999).
These are impressive dynamical phenomena that we
happen to be able to observe with a certain degree of 
completeness in both white light (Dere et al. 1999,
St. Cyr et al. 2000, Burkepile et al. 2004)
as well as in X-ray and EUV bands (Webb \& Hundhausen 
1987, Hudson \& Cliver 2001). In a low-$\beta$ plasma
system on the large scales of the corona, we generally
believe that MHD processes control the dynamical
evolution of the magnetic field and the coronal
plasma. In considering CMEs as a MHD phenomena,
previous studies have mainly focused on two physical issues:
magnetic energy storage and CME initiation.

The energy storage problem is concerned with how much
free magnetic energy can be stored in the corona in
order for the field to spontaneously open up to infinity 
during a CME eruption. This issue was first brought up in
Aly (1984), and later on in Aly (1991) and Sturrock (1991), 
where Aly conjectured that a fully-anchored force-free
field may not be able to store enough free magnetic energy 
to drive a CME kinematically, in addition to opening up the 
field under conditions of high electrical conductivity.
This conjecture states that an anchored force-free field 
cannot store more energy than that of any fully-open field
sharing the same boundary flux distribution at the 
base of the atmosphere, where those fully-open fields have
energies bounded from below by a minimum usually referred 
to as the Aly limit. Many studies have pointed out that
the constraint of the Aly Conjecture can be bypassed
by relieving one of the conditions assumed in the 
Conjecture. For example, by relieving the force-free
condition an equilibrium field with plasma weight can
easily store enough free magnetic energy to drive the CME
(Low \& Smith 1993, Wolfson \& Dlamini 1997, Fong et al. 2002, 
Low et al. 2003, Zhang \& Low 2004, Flyer et al. 2005).
Even under the force-free condition, recent studies have
showed that the Aly limit can be exceeded if the field 
contains a detached magnetic flux rope in the corona (Hu et al. 
2003, Wolfson 2003, Flyer et al. 2004). The Aly Conjecture
can also be bypassed through the topology of multipolar fields.
For multipolar fields, a CME-like expulsion only needs to open 
up one but not all the bipolar fields (Antiochos et al. 1999).
In such a situation, the Aly limit associated with opening up
the entire field is not the relevant energy threshold; see
Zhang \& Low (2001) for illustrative examples.

The CME initiation problem is concerned with the sporadic
behavior and locations of CME occurrences. CMEs are occurring 
at an average rate of one to three events per day, with the 
frequency of occurrence increasing as solar activity 
intensifies. The catastrophe model (Forbes \& Isenberg 1991,
Isenberg, Forbes \& Demoulin 1993) interprets the CME
initiation as a catastrophic loss of equilibrium in a series 
of quasi-static evolution of magnetic equilibria corresponding 
to mathematical variations of some model parameters.
The flux-emergence trigger model (Chen \& Shibata 2000)
proposes that the newly-emerged flux near a preexisting 
flux rope is the trigger of the expulsion.

It might be useful to point out that when we address the 
magnetic energy storage problem or the CME initiation problem, 
our underlying thinking tends to take these as two independent 
issues. However, as we have pointed out in a recent review 
(Zhang \& Low 2005) these two issues can be related by
the accumulation of magnetic helicity in the corona as
a fundamental MHD process. The accumulation of magnetic
helicity in the corona leads to a magnetic energy
build-up as a natural product of coronal evolution.
Eruptions, as a loss of field confinement, occur when 
the stored magnetic energy exceeds the Aly limit.
In this paper, we take another approach to address
the role of magnetic helicity accumulation in the corona.
We propose a hydromagnetic conjecture that there is an 
upper bound on the total magnetic helicity that a 
force-free field can contain, owing to the fact that a 
force-free field in an open atmosphere has to self-confine.
This conjecture poses a fundamental but mathematically 
challenging problem in magnetohydrodynamics.
We focus our attention in this paper on demonstrating
the physical plausibility of the conjecture by using
several families of numerical solutions of nonlinear 
force-free fields to make our physical points.

The paper is organized as follows:
In \S 2, the physical considerations on the roles of magnetic
helicity accumulation are addressed, with a particular focus
on their connections to the magnetic energy storage and
CME initiation. In \S 3, power-law axisymmetric force-free
fields are used to indicate the possible existence of
an upper bound on the total magnetic helicity, using 
both analytical inequalities and numerical solutions. 
Conclusions and discussions are presented in \S 4.


\section{The Physical Picture and the Conjecture}

\subsection{Magnetic helicity and its approximate
conservation in the corona}

Magnetic helicity is defined as 
$H(V) = \int_V {\bf \omega} \cdot {\bf B} dV$,
where ${\bf \omega}$ is the vector potential of the 
vector magnetic field ${\bf B}$ in a volume $V$.
It is a physical quantity that quantitatively measures the
topological complexity of a magnetic field, such as
the degree of linkage and/or twistedness in the field,
provided $V$ wholly contains ${\bf B}$.  We henceforth
refer to $V$ as a a magnetic volume.
With this physical quantity, the frozen-in condition
in a perfectly conducting plasma can then be stated 
in terms of the absolute conservation of magnetic
helicity in all magnetic sub-volumes as well as of the 
total magnetic helicity for the whole magnetic volume
(Moffatt 1985, see Zhang \& Low 2005 for more descriptions).

In a situation of a domain $V$ whose magnetic field ${\bf B}$ 
threads across its boundary, i.e., the normal magnetic field 
$B_n$ does not vanish across the domain boundary $\partial V$, 
$H(V)$ is then gauge dependent and is therefore not physical.
Magnetic twist and linkage in $V$ then must be described by 
the relative magnetic helicity denoted by $H_R(V)$
introduced by Berger \& Field (1984). We will define 
$H_R(V)$ later with the development of the paper.
In this paper, we take magnetic helicity to mean
either $H(V)$ or $H_R(V)$, depending on whichever applies 
to the given situation.

The frozen-in condition, or the absolute conservation
of magnetic helicity in magnetic volumes, must break down
in a realistic plasma where the electrical conductivity is 
very high but not infinite. Electric current sheets can form, 
and magnetic reconnection will subsequently set in, to
dissipate the thin current sheets and change field 
topology (Parker 1994). However, by a general scaling 
analysis of resistive processes, Berger (1984)
pointed out that the dissipation rate of the total relative 
magnetic helicity in the corona is orders of magnitude 
smaller than the release rate of magnetic energy. 
So the total magnetic helicity in the corona is still 
approximately conserved despite the occurrences
of magnetic reconnection.

The conservation of total magnetic helicity in the
corona has a profound influence on the coronal dynamics.
Magnetic helicity is continuously being transported into 
the corona (Kurokawa 1987, Lites et al. 1995, Leka et al. 
1996, Zhang 2001, Pevtsov, Maleev \& Longcope 2003) with
a preferred sign for the two solar hemispheres, positive
and negative in the southern and northern hemisphere,
respectively (Rust 1994, Pevtsov, Canfield \& Metcalf 1995, 
Bao \& Zhang 1998). Moreover, the same opposite signs of helicity
are statistically preferred in the two respective hemispheres,
negative and positive, respectively, in the northern and 
southern hemispheres, independent of the solar cycle. 
Thus, under the approximate conservation of total magnetic 
helicity, the hemispherical preferred signs of helicity
implies an accumulation of magnetic helicity in the two 
hemispheres. It is this accumulated helicity that, we 
propose, characterizes the build up of magnetic-energy
and when accumulated to a point a magnetic field must open 
up as an eruption.


\subsection{Helicity-energy relationship for fields in finite domains}

For a force-free field confined in a finite domain with
rigid walls, its total magnetic energy can exceed any value 
by endowing the field with a sufficiently twisted topology.
This can be easily seen from the Virial Theorem of 
Chandrasekhar (1961), which states that for a force-free
field confined in a sphere of radius $r = 1$, its total 
magnetic energy, given in spherical coordinates, is
\begin{equation}
E_0  =  \int_{r < 1} {B^2 \over 8 \pi} dV
        =  \frac{1}{4} \int_{r = 1}
	\left( B_{\theta}^2 + B_{\varphi}^2 - B_r^2
        \right) \sin \theta d\theta ~~.
\end{equation}
\noindent
For a fixed boundary flux $B_r$ at $r = 1$, the more highly
twisted the force-free field is, the greater are the boundary 
values of the tangential components $B_{\theta}$ and 
$B_{\varphi}$ at $r = 1$, leading to a higher total
magnetic energy of the field. Note that force-free solutions
can exist in such a domain even if $B_r = 0$ on $r = 1$.

More importantly, if the field is given a total non-zero 
magnetic helicity $H_0$, Woltjer (1958) Theorem sets 
a ground-level energy above that of the potential field.
The Woltjer Theorem produces
many constant $\alpha$ force-free fields, each being a local
minimum-energy state compared to neighboring fields with 
the same total magnetic helicity $H_0$. The absolutely lowest
minimum-energy state, taken from among all those constant
$\alpha$ force-free fields, is called the Woltjer state, 
characterized by a specific constant $\alpha_0$. This constant
$\alpha_0$ of the Woltjer state, together with the given
total magnetic helicity $H_0$, puts a bound from below 
for the magnetic energy of the field in a finite domain.
That is, the magnetic energy of any field ${\bf B}$
possessing the same total magnetic helicity of $H_0$ 
satisfies the bound:
\begin{eqnarray}
E > | ~ \alpha_0 H_0 ~ | ~~.
\end{eqnarray}

Applying this theorem to a coronal magnetic field underlying 
a locally confined coronal structure (Zhang \& Low 2003),
we can see that the coronal field cannot relax to a 
potential state as long as the total
magnetic helicity is not zero. Magnetic reconnection 
may set in to release the excess magnetic energy that 
the coronal field may happen to have, but the field cannot 
relax to a state that possesses less magnetic energy than 
that of the Woltjer state.
With more and more magnetic helicity brought with a preferred
sign into 
the corona, this ground-level magnetic energy increases with 
the increasingly accumulated helicity. This is the principal
consideration described in Zhang \& Low (2005) of
the role of the accumulated magnetic helicity in storing
magnetic energy for CMEs.


\subsection{Confinement issue for fields in unbounded domain}

For a force-free field filling an unbounded atmosphere,
it has an energy property that is distinct from the one
confined in a finite domain with rigid walls. Now, the
Virial Theorem of Chandrasekhar (1961) takes a different
form and sets a special condition for those fields
in force-free equilibrium. For a force-free field outside
a sphere of radius $r = 1$ in spherical coordinates,
its total magnetic energy now takes the form of
\begin{equation}
E_0  =   \int_{r > 1} {B^2 \over 8 \pi} dV
     = \frac{1}{4} \int_{r = 1}
	\left( B_r^2  - B_{\theta}^2 - B_{\varphi}^2
        \right) \sin \theta d\theta ~~,
\end{equation}
\noindent
where we assume that $|{\bf B}|$ spatially decline to zero
at $r \rightarrow \infty$ so that the total magnetic energy
$E_0$ is finite. Here and elsewhere we assume this is the 
case whenever we treat integration over unbounded domain
without explicitly stating this qualification.
Since $E_0$ should always be positive, the above equation 
says that no force-free state is possible in $r > 1$ domain 
unless $B_r \ne 0$ on $r = 1$. With $B_r \ne 0$, equation (3)
sets an absolute upper bound on the total magnetic energy,
\begin{equation}
E_{max} = \frac{1}{4} \int_{r = 1} 
B_r^2 \sin \theta d\theta ~,
\end{equation}
\noindent which is uniquely determined by the flux
distribution on $r = 1$, independent of the twist of the
field.

The essence of this hydromagnetic result is that there is
an upper bound on the total magnetic energy that a force-free
field can attain and this upper bound is determined by its
boundary flux at $r = 1$. This is equivalent to saying 
that a force-free field in an unbounded atmosphere 
requires certain means of confinement to remain in 
equilibrium. The field with with an arbitrarily large, 
prescribed degree of twist may not be able to find a 
force-free state to meet the bound on its total magnetic energy.
This is indicated by the the following analysis.

A field completely detached from $r = 1$,
as drawn in Figure 1b where $B_r \equiv 0$, can never be
in equilibrium in the absence of other body forces.
The only way to hold a flux rope, and hence attain
a certain amount of magnetic helicity in the low corona, 
is to add a $B_r \ne 0$ field as in Figure 1a to
function as an anchoring agent. 
The resultant configuration will be the one sketched in
Figure 1c, a field containing a certain amount of
magnetic helicity and yet finding equilibrium in the
$r > 1$ domain.

Now imagine we fix the $B_r$ distribution on $r = 1$ in
Figure 1c but arbitrarily increase the total azimuthal flux 
in the flux rope. This parametric change will take the field
closer and closer to the configuration in Figure 1b.  We 
intuitively expect that there would be a parameter 
representing the
field topology crossing over where no force-free equilibrium
can exist. We suggest that this critical parameter corresponds
to a point of maximum accumulation of magnetic helicity 
in the atmosphere.

This conjecture intuitively comes from the understanding 
that the total magnetic helicity cannot be destroyed 
in the corona at observable time scales. So the corona
may thus get a chance to accumulate magnetic helicity of a
preferred sign to cross over a threshold bounding the 
allowable total helicity in a force-free field.  A rigorous 
mathematical bound on the total helicity of a force-free field
in $r > 1$, in full generality, is needed to prove the 
conjecture, in parallel to equation (3) for the bound on the 
total energy of a force-free field. This is a difficult
task in mathematical physics. As a first step towards such
a proof, we use families of power-law axisymmetric
force-free fields of Flyer et al. (2004)
to investigate the nature of this conjecture. 
We focus the rest of our paper on such a study in order to
provide a basic understanding of the physical issues.


\section{Helicity Bounds for Power-law Axisymmetric Force-free Fields}

\subsection{A Formula of Total Relative Magnetic
Helicity of Axisymmetric Fields}

We first derive a formula to calculate the
total relative magnetic helicity in axisymmetric magnetic
fields. This simple formula is limited to axisymmetric
fields but is otherwise quite general, independent of
whether the axisymmetric field is in equilibrium or not.
We therefore derive the formula from first principles.

Under axisymmetry, we can always
write the solenoidal magnetic field ${\bf B}$ in $r > 1$
in the form of
\begin{equation}
{\bf B} = {1 \over r \sin \theta} \left[
{1 \over r}{\partial A \over \partial \theta} ~,
~ - {\partial A \over \partial r}~, ~Q (A) \right] ~.
\end{equation}
\noindent
Here, the flux function $A$ defines the poloidal magnetic
field and the function $Q$ defines the toroidal field.

Alternatively, we can also use a vector potential $\omega$
to define a solenoidal magnetic field, where the magnetic
field ${\bf B}$ can be derived from the vector potential 
$\omega$ by the following relationship:
\begin{equation}
{\bf B} = \nabla \times \omega ~~.
\end{equation}
\noindent Then for an axisymmetric magnetic field,
$\omega$ takes a form of
\begin{equation}
{\bf \omega} = {1 \over r \sin \theta} \left[
{1 \over r}{\partial Q' \over \partial \theta} ,
- {\partial Q' \over \partial r}, A' \right],
\end{equation}
\noindent
where $Q'$ and $A'$ are two physical quantities that are
related to the functions $A$ and $Q$ by
\begin{equation}
L Q'  \equiv   {\partial^2 Q' \over \partial r^2}
+ {1-\mu^2 \over r^2}
{\partial^2 Q' \over \partial \mu^2} = - Q ~~,
\end{equation}
\noindent
with $\mu = \cos \theta$, and
\begin{eqnarray}
A'   =  A + constant  ~~.
\end{eqnarray}

These relationships between $Q'$, $A'$, $Q$ and $A$ can be understood
by the following analysis. From the definition of the vector 
potential and Stokes Theorem, the magnetic flux across any 
surface $S$ with a boundary $C$ is $ F = \int_S {\bf B} 
\cdot d{\bf S} = \int_C {\omega} \cdot d{\bf l} $
where $d{\bf l}$ is a path element along $C$. Therefore,
the vector potential $\omega$ carries information about 
the magnetic flux across surfaces. 
Application of this relation with $\omega$ given by
equation (7) for an axisymmetric field shows that the function
$Q'$ carries information on the azimuthal flux, as implied by
the relationship between $Q'$ and $Q$ in equation (8), and the
function $A'$ carries information about the poloidal flux 
as expressed by equation (9).

The $constant$ in equation (9) can be used to take
care of the physical requirement that $A'$ should be zero
along the polar axes.
Without loss of generality, we set this $constant$ to
zero, which is equivalent to setting the flux function $A$
to zero along the polar axes.

Suppose we are given the field ${\bf B}$ in the unbounded
space $r > 1$ denoted as $V$.  This field has a flux
anchored to the inner boundary $r = 1$ so that the 
classical helicity of Woltjer is not gauge invariant
and we must use instead the relative helicity of 
Berger \& Field (1984).  This is given by the formula
derived in Berger (1985):
\begin{equation}
H_R = \int _{V_a} \omega^p_a \cdot {\bf B}^p_a \hspace{2mm} dV
+  \int _{V} \omega \cdot {\bf B} \hspace{2mm} dV ~ ,
\end{equation}
\noindent
which involves extending the domain $V$ to the spherical
region $r < 1$ denoted as $V_a$ where ${\bf B}_a^p$ is
a potential field with vector potential $\omega_a^p$. 
This potential field continues into the given
field ${\bf B}$ across the boundary with continuity of $B_r$.
The important point here is 
that $H_R$ is independent of the gauge and is a physical
quantity describing the topological complexity of the field 
${\bf B}$ in $r > 1$. We calculate $H_R$ as follows.

The solenoidal condition requires
\begin{equation}
{\bf B}_a^p \cdot {\bf r} \hspace{2mm} |_{r=1} =
{\bf B} \cdot {\bf r} \hspace{2mm} |_{r=1}
\end{equation}
\noindent 
for the continuity of the normal field across $r = 1$.
Equivalently, in terms of the vector potential, we require
\begin{equation}
\omega_a^p \times {\bf r} \hspace{2mm} |_{r=1} =
\omega \times {\bf r} \hspace{2mm} |_{r=1} ~ ,
\end{equation}
\noindent
for the continuity of the tangential components of the vector
potential (Berger \& Field 1984).

Equation (12) requires that
both $A$ and ${\partial Q' \over \partial r}$ are continuous
across $r = 1$. The continuity of $A$ across $r = 1$ is just
equation (11) and the normal flux of ${\bf B}$ at $r = 1$
determines a unique potential field ${\bf B}^p_a$ in $V_a$.
This potential field is generated by a flux function $A_a^p$, 
with $Q = 0$. We take $Q' = 0$ everywhere in $V_a$, which is
consistent with equation (8) where $Q = 0$, as a selected gauge. 
Hence, we have
\begin{eqnarray}
{\bf B}^p_a & = & {1 \over r \sin \theta}
\left( {1 \over r} {\partial A_a^p \over \partial \theta}{\hat r}
 - {\partial A_a^p \over \partial r} {\hat \theta} \right) \nonumber \\
\omega^p_a & = & {A \over r \sin \theta} {\hat \varphi},
\end{eqnarray}
\noindent
which are two orthogonal vectors so that 
$\int_{V_a} \omega_a^p \cdot {\bf B}_a^p dV = 0$ and
$H_R$ in equation (10) then contains only the contribution 
from $r > 1$.

Now we have the total relative magnetic helicity in $r > 1$ as
\begin{eqnarray}
H_R  =  \int_{r>1} {\bf \omega} \cdot {\bf B} dV
  =  2 \pi \int_{r>1} {1 \over r^2 (1-\mu^2)}\left[
   \nabla Q' \cdot \nabla A + A Q \right] r^2 dr d\mu ~~,
\end{eqnarray}
\noindent
where we have the geometric boundary conditions that both
$Q'$ and $A$ vanish at the polar axes and the gradients
of these two functions with distance vanish sufficiently
fast at infinity.
These conditions and integrations by parts allow us to
rewrite
\begin{eqnarray}
\int_{r>1} {1 \over r^2 (1-\mu^2)}
   \nabla Q' \cdot \nabla A  r^2 dr d\mu
   = \int_{r>1} dr d\mu {AQ \over 1-\mu^2}
   - \int_{-1}^1 {d\mu \over 1-\mu^2}
 \left[A {\partial Q' \over \partial r} \right]_{r = 1} ~.
\end{eqnarray}
\noindent
And then the total relative magnetic helicity is given by:
\begin{eqnarray}
\label{helicity}
H_R & = & 4 \pi \int_{r>1} dr d\mu {AQ \over 1-\mu^2}
 - 2 \pi \int_{-1}^1 {d\mu \over 1-\mu^2}
    \left[A {\partial Q' \over \partial r} \right]_{r = 1} \nonumber \\
 &  = & 4 \pi \int_{r>1} dr d\mu {A Q \over 1-\mu^2} ~~,
\end{eqnarray}
\noindent
a simple form of an integral to calculate the total
relative magnetic helicity in $r > 1$.

Note this formula can also be rewritten as
\begin{equation}
H_R = 4 \pi \int_{r>1}  A \times B_{\varphi} r dr d \theta
\end{equation}
\noindent
and it gives a physical meaning of this simple form of
helicity formula: The total relative magnetic
helicity in axisymmetric fields is just a simple convolution
of the local wrapping of the azimuthal flux element
$B_{\varphi} r dr d \theta$
with the poloidal flux $A$.


\subsection{Power-law axisymmetric force-free fields}

We adopt the families of numerical force-free fields from
Flyer et al. (2004) where $Q$ as a strict function of $A$ 
has the form of
\begin{eqnarray}
Q^2 (A) = \frac{2 \gamma}{n+1} A^{n+1} ~.
\end{eqnarray}
\noindent
Here $n$ is an odd constant index required to be no less
than 5 in order for the field to possess finite magnetic
energy in $r > 1$ and $\gamma$ is a free parameter which we
choose to be positive without loss of generality.
We refer this family of force-free fields
power-law axisymmetric force-free fields in this paper.

Taking this form of $Q$, the force-free condition, that is,
\begin{eqnarray}
\left(\nabla \times {\bf B}\right)  \times {\bf B} = 0 ~~,
\end{eqnarray}
\noindent reduces to the following governing equation
for flux function $A$:
\begin{eqnarray}
{\partial^2 A \over \partial r^2} + {1 - \mu^2 \over r^2}
{\partial^2 A \over \partial \mu^2} + \gamma A^n = 0 ~~.
\end{eqnarray}
\noindent
This governing equation was solved numerically as a boundary
value problem within domain $r > 1$ in Flyer et al. (2004),
subject to the prescribed boundary flux distribution
\begin{equation}
\label{bc}
A|_{r = 1} = \sin^2 \theta ~~.
\end{equation}
\noindent
We refer interested readers to that paper
for various properties of this family of force-free fields
for the cases of $n = 5, 7, 9$.

Note that although $\gamma > 0$ is a free parameter,
it actually has a least upper bound $\gamma_0$ above 
which no solution of the boundary value problem can exist.
As shown in Flyer et al. (2004), with the boundary 
condition (\ref{bc}), for each fixed $n$,
\begin{equation}
\gamma < \gamma_0 = {\frac{2(n+1)(2n+1)!! }{3 (2n)!!}} ~.
\end{equation}


\subsection{A bound on the total magnetic helicity
for a fixed $n$}

The Virial Theorem expresses the total magnetic energy of 
a force-free magnetic field ${\bf B}$ in $r > 1$ in terms 
of a surface integral given by equation (3). 
Aly (1988) presented a ``generalized scalar
virial equation'' that expresses the total magnetic energy in 
other forms. Writing $B_t^2 = B_{\theta}^2 + B_{\varphi}^2$
and introducing an arbitrary function 
$f(r)$ with derivative $f^{\prime}(r)$, Aly derived
\begin{eqnarray}
\int_{r>1} \left[ f (B_r^2+B_t^2) +
 r f^{\prime} (B_t^2-B_r^2) \right] dV ~~
        = ~~ f(r=1) \int_{r = 1}  (B_r^2  - B_t^2) dS ~~.
\end{eqnarray}
\noindent 
For our purpose, we set $f = r$ to obtain
\begin{eqnarray}
\int_{r>1} \left[ r (B_r^2+B_t^2) +
 r (B_t^2-B_r^2) \right] dV
        = \int_{r = 1}  (B_r^2  - B_t^2) dS ~~.
\end{eqnarray}
\noindent
This leads to an alternative expression of total magnetic
energy of force-free fields as
\begin{eqnarray}
E_0 &   =  & \frac{1}{4 \pi} \int_{r>1} r \left(B_{\theta}^2
          + B_{\varphi}^2 \right) dV ~~.
\end{eqnarray}
\noindent
From this form of $E_0$, we then have
\begin{eqnarray}
E_0   & \ge & \frac{1}{4 \pi} \int_{r > 1} r B_{\theta}^2 dV \\
E_0   & \ge & \frac{1}{4 \pi} \int_{r > 1} r B_{\varphi}^2 dV
\end{eqnarray}
\noindent and
\begin{eqnarray}
E_0^2 & \ge &  \int_{r > 1} \frac{r B_{\theta}^2}{4 \pi} dV
\int_{r > 1} \frac{r B_{\varphi}^2}{4 \pi} dV ~~.
\end{eqnarray}

Consider Cauchy-Schwartz inequality
\begin{equation}
\int_{X} f^2 dx \int_{X} g^2 dx 
\ge \left(\int_{X} fg dx \right)^2 ~~,
\end{equation}
\noindent  
which relates two functions $f$ and $g$ in space $X$,
where $dx$ is the spatial integration element.
Applying Cauchy-Schwartz inequality to inequality (28), 
we get
\begin{eqnarray}
E_0^2 & \ge &  \left(\frac{1}{4 \pi} \int_{r > 1}
 r B_{\theta} B_{\varphi} dV \right)^2  ~~~.
\end{eqnarray}
\noindent
The right hand side of this inequality contains an integral 
of the product of the poloidal field $B_{\theta}$ with the 
toroidal field $B_{\varphi}$, suggestive of the formula (16) 
of the relative magnetic helicity.

By equation (16) the total relative magnetic helicity 
in $r > 1$ of a power-law axisymmetric force-free field
is
\begin{eqnarray}
H_R & = & 4 \pi \sqrt{\frac{2 \gamma}{n+1}}
  \int_{-1}^1 {d\mu \over 1-\mu^2}
  \int_{1}^{\infty} A^{\frac{n+3}{2}} dr ~~.
\end{eqnarray}
Evaluate the right hand side of inequality (30) we obtain
\begin{eqnarray}
 \frac{1}{4 \pi} \int_{r > 1} r B_{\theta} B_{\varphi} dV 
 & = & - \frac{1}{2} \int_{-1}^1  {d\mu \over 1-\mu^2}
    \int_{1}^{\infty} r dr {\partial A \over \partial r}
 \sqrt{\frac{2 \gamma}{n+1}} A^{\frac{n+1}{2}} \nonumber \\
 & = & \frac{1}{2} \int_{-1}^1 {d\mu \over 1-\mu^2}
  [{2 \over n+3} \sqrt{\frac{2 \gamma}{n+1}} 
  A^{\frac{n+3}{2}}]_{r=1} \nonumber  \\ &
  & +\frac{1}{2} \int_{-1}^1 {d\mu \over 1-\mu^2}
     \int_{1}^{\infty} dr {2 \over n+3} 
     \sqrt{\frac{2 \gamma}{n+1}} A^{\frac{n+3}{2}} ~~,
\end{eqnarray}
\noindent with an integration by part.
Using the boundary condition at $r = 1$ to evaluate the 
first term and denoting it as
${2 \over n+3} \sqrt{\frac{2 \gamma}{n+1}} c_0$, then
\begin{equation}
c_0 = \frac{1}{2} \int_{-1}^1 {d\mu \over 1-\mu^2}
 [A^{\frac{n+3}{2}}]_{r=1} > 0 
\end{equation}
\noindent is a constant independent of $\gamma$. 
Inequality (30) now takes the form of
\begin{eqnarray}
(\frac{n+3}{2})^2 E_0^2 & \ge
   &  \left(\sqrt{\frac{2 \gamma}{n+1}} c_0
   + \frac{H_R}{8 \pi} \right)^2  ~~,
\end{eqnarray}
and we have achieved a bound on $H_R$.

A simple analysis reduces inequality (34) to the form of
\begin{equation}
\label{inequality12}
-8 \pi \left[ {n+3 \over 2} E_0 +
\sqrt{{2 \gamma \over n+1}} c_0 \right]
\le H_R \le 8 \pi \left[{n+3 \over 2} E_0
- \sqrt{{2 \gamma \over n+1}} c_0 \right] .
\end{equation}
\noindent 
Since $c_0$ is defined strictly by the boundary condition,
inequality (\ref{inequality12}) shows that the magnitude
of $H_R$ is bounded from above by bounds that are fixed
for each given $n$.

The above inequality shows that $|H_R|$ is bounded for 
each fixed $n$. However, this does not assure us
that a single bound $H_{max} > 0$ exists such that 
$|H_R| < H_{max}$ for all $n$, because the upper and
lower bounds go to $\pm \infty$ as $n \rightarrow \infty$.
A tighter bound on $H_R$ is then motivated to be found and
will be presented in \S 3.5.


\subsection{Numerical evidence}

It is useful for us to use those numerical solutions in
Flyer et al. (2004) to examine some helicity properties of
this family of power-law axisymmetric force-free fields.
In Flyer et al. (2004), Newton's iteration
combined with a pseudo-arc length continuation scheme
was used to solve the nonlinear partial differential 
equation (20) in the unbounded domain $r > 1$ for the 
boundary condition (21). It was found that iterations 
converge to solutions only for a finite range of $\gamma$ 
for a fixed index $n$, as indicated by inequality (22). 
In addition, the solution $A$ was found to be a multi-valued 
function of $\gamma$, i.e. for a given value of $\gamma$ 
multiple solutions can exist. These multiple functions can 
be pieced into a single solution curve as a function of 
$\gamma$ joined at turning points. The pseudo-arc length 
continuation scheme was used to march along the solution 
curve for a given $n$ and to pass through the turning points 
in the parameter space where the Jacobian of the Newton's 
iteration become singular.

Each point on the obtained solution curves describes a
power-law axisymmetric force-free field in $r > 1$ 
with a total magnetic energy $E_0$, a total
azimuthal flux
\begin{equation}
F_{\varphi} = \int_{r > 1} B_{\varphi} r dr d\theta ~,
\end{equation}
\noindent
and a total relative magnetic helicity $H_R$ given by 
equation (16). Flyer et al. (2004) characterized
their solutions in terms of $E_0$ and $F_{\varphi}$, 
but not $H_R$. The approximate conservation law for $H_R$ 
naturally suggests the use of $H_R$ to characterize the 
twisted state of the force-free field in $r > 1$. In
Figure 2 we display the variation of $E_0$ and $F_{\varphi}$
with $\gamma$ for the $n = 5, 7, 9$ solution curves, 
taken from Flyer et al. (2004), together with the 
respective variations of $H_R$ for these curves.

For each of the $n = 5, 7, 9$ solution curves, 
Flyer et al. (2004) found that $F_{\varphi}$ is bounded 
to be of the order of twice the total poloidal flux 
given by boundary
condition (\ref{bc}). This numerical evidence indicates
that force-free fields can only self-confine
its magnetic pressure by its tension force provided its
azimuthal flux is less than a bound determined by its
poloidal flux fixed by the boundary condition.
These are indicated in Figure 2 by the azimuthal flux 
$F_{\varphi}$ monotonically increases from 0 to its maximum 
value along a solution curve, noting in particular that
the maximum value of $F_{\varphi}$ is in each case less 
than 2 and not sensitive to $n$. The corresponding relative 
helicity $H_R$ shows interesting undulations superposed on 
a general trend of monotonic increase along the solution curves,
showing a bound on $H_R$ less than 15 and not sensitive to $n$.
These numerical solutions provide a basis for the suggestion 
that the total relative magnetic helicity of the three families 
of power-law axisymmetric force-free fields is also bounded
by a suitably defined bound that depends only on the 
boundary flux distribution.

Figure 3 presents these solutions again, but are plotted for
the total magnetic helicity (left panels) or for the total
magnetic helicity divided by the total magnetic energy
(right panels) against the azimuthal flux.
These plots show that the total magnetic helicity
increases nearly monotonically with the increasing
azimuthal flux, which confirms our intuition that the
confinement issue discussed in terms of the azimuthal flux
is not much different from the discussion in terms of
magnetic helicity. However, the total magnetic helicity
is known to be conserved in the corona and therefore
accumulated in the corona, whereas the physics is less
specific whether the azimuthal flux is conserved 
in the corona or not.

Another interesting observation from Figure 3 is that, the
ratio between the total magnetic helicity and the total
magnetic energy shows a better monotonic variation with the
azimuthal flux than does the magnetic helicity itself.
This may reflect a general relationship among the three
quantities. Notice that inequality (35) suggests
that $H_R/E_0$ is roughly smaller than $4\pi(n+3)$ which is
confirmed by the numerical solutions. However,
the bound obeyed by the numerical solutions is a lower
or tighter upper bound. That is,
$H_R/E_0 < 9\pi < 4\pi(n+3)=(32-48)\pi$, the last term taken
from inequality (35). This hints that the upper bound we have
found by inequality (35) is not a stringent bound and that 
lower upper bounds may exist.


\subsection{Inequality relating total magnetic helicity
to azimuthal flux and magnetic energy}

The numerical relationships among $E_0$, $F_{\varphi}$
and $H_R$ in Figure 3 suggest that a rigorous relationship
may exist between these three physical quantities, 
at least, for the power-law force-free fields.
This rigorous relationship is derived in this subsection
by first introducing
\begin{equation}
E_{\varphi} = \frac{1}{8\pi} \int_{r > 1} B_{\varphi}^2 dV ~,
\end{equation}
\noindent
which is the contribution of the azimuthal field component
to the total magnetic energy.

As an extension of Cauchy-Schwartz inequality,
Holder inequality states that, given any two functions 
$f$ and $g$ in space $X$ with spatial integration 
element $dx$, we have
\begin{equation}
\label{holder}
\left(\int_{X} f^k dx \right)^{1/k}\left( \int_{X} g^{k'} dx
\right)^{1/k'} \ge \int_{X} fg dx  ~,
\end{equation}
\noindent
where $k > 1$ and is related to $k'$ by
\begin{equation}
\label{kk'}
\frac{1}{k} + \frac{1}{k'} = 1 ~~.
\end{equation}
\noindent
The inequality (\ref{holder}) reverses direction if $k < 1$ 
while k still relates to $k'$ by equation (\ref{kk'}).

Applying Holder inequality with
\begin{eqnarray}
f & = & A^2 , \\
g & = & A^m , \\
k   & = & m+1 , \\
k'  & = & 1 + 1/m ,\\
dx & = & {dr d\theta \over \sin \theta} ,
\end{eqnarray}
\noindent we have
\begin{equation}
\label{inequlity3}
\left(\int_{r > 1} A^{2m+2} {dr d\theta \over \sin \theta} 
\right)^{\frac{1}{m+1}}
\left(\int_{r > 1} A^{m+1} {dr d\theta \over \sin \theta} 
\right)^{\frac{m}{m+1}}
\ge \int_{r > 1} A^{m+2} {dr d\theta \over \sin \theta} ~,
\end{equation}
\noindent
where $m=\frac{n-1}{2}$ and is an integer greater than 1.
With the power-law definition of $Q$ in equation (18), 
it is easy to show that
\begin{eqnarray}
E_{\varphi} & = & \frac{\gamma}{4(m+1)}
\int_{r > 1} A^{2m+2}
{dr d\theta \over \sin \theta} ~~, \\
F_{\varphi} & = & \sqrt{\frac{\gamma}{m+1}}
\int_{r > 1} A^{m+1} {dr d\theta \over \sin \theta} ~~, \\
H_R         & = & 4 \pi \sqrt{\frac{\gamma}{m+1}}
\int_{r > 1} A^{m+2} {dr d\theta \over \sin \theta} ~~.
\end{eqnarray}
\noindent
Now inequality (\ref{inequlity3}) leads to
\begin{equation}
H_R \le 4 \pi F_{\varphi}^{\frac{m}{m+1}}
          E_{\varphi}^{\frac{1}{m+1}}
     \left[\frac{16(m+1)}{\gamma}\right]^{\frac{1}{2m+2}} ~.
\end{equation}

Define
\begin{eqnarray}
y & = & \frac{H_R}{4 \pi F_{\varphi}^{\frac{m}{m+1}}
          E_{\varphi}^{\frac{1}{m+1}}
     \left[\frac{16(m+1)}{\gamma}\right]^{\frac{1}{2m+2}}} ~.
\end{eqnarray}
\noindent
Inequality (49) is then equivalent to say that $y \le 1$.

Figure 4 presents the $y$ values, plotted against
$F_{\varphi}$, for the families of $n=5, 7, 9$ fields, 
using numerical solutions in Flyer et al. (2004) again.
We see that the y values for these three families of fields
vary between 0.75 to 0.95, with a tendency for the overall y
values to increase with the $n$ values. The fact that these y
values are all less than 1 is just the verification of
inequality (49), which is rigorous
for all power-law axisymmetric force-free fields. 
The more interesting point is that these y values
can get close to 1. This implies that we are getting
close to finding a stringent bound on the total magnetic
helicity by inequality (49), a more stringent inequality
than inequality (35) is.

As pointed out by the referee, an interesting feature from
Figure 4 is that our y values do not vanish to zero when
both the total magnetic helicity and azimuthal flux vanish.
Inserting equations (46)-(48) into equation (50), we find
that our y values can be represented as 
\begin{equation}
y  =  \frac{\int_{r > 1} A^{m+2} {dr d\theta \over \sin \theta}}
    {(\int_{r > 1} A^{m+1} {dr d\theta \over \sin \theta})^{\frac{m}{m+1}}
 (\int_{r>1}A^{2m+2}{dr d\theta \over \sin \theta})^{\frac{1}{m+1}}} ~,
\end{equation}
\noindent
which shows that values of y are not directly dependent on $\gamma$.
Applying potential field $A=\frac{\sin^2\theta}{r}$, we get
$ y = 0.88, 0.90, 0.92 $ for $n = 5, 7, 9$ respectively.
These numbers are consistent with the tendencies of y values 
in Figure 4.

Now multiply across equation (20) with flux function $A$
and then integrate it with the element
${dr d\theta \over \sin\theta}$ to get
\begin{equation}
\int_{r > 1} (m+2) Q^2 dr {d\theta \over \sin \theta} =
\int \left[A {\partial A \over \partial r}
\right]_{r = 1}{d\theta \over \sin \theta} +
\int_{r > 1} \left(B_r^2 + B_{\theta}^2
 + B_{\varphi}^2 \right)
dr {d\theta \over \sin \theta} ~.
\end{equation}
\noindent
The left-hand side of the equation is proportional 
to $E_{\varphi}$, whereas, on the right hand
side, the first term is bounded by the Chandrasekhar
Virial Theorem and the second term is proportional to the 
magnetic energy $E_0$.  Direct derivation then gives
\begin{equation}
(m+2) E_{\varphi} \le \frac{4}{3} E_{max}+ E_0 ~~,
\end{equation}
\noindent
where $E_{max}$ is given by equation (4).

As $E_0<E_{max}$ and $E_{max}$ is bounded
by the boundary condition at $r=1$, the above inequality (53)
shows that $(m+1)E_{\varphi}$ is also bounded. In consequence,
we have $E_{\varphi} \rightarrow 0$ 
as the free index $m \rightarrow \infty$.
This means that as we go to higher and higher $n$,
the energy contributed by $B_{\varphi}^2$ becomes negligible.
Much of the energy in the sum $B_{\theta}^2 + B_{\varphi}^2$
is in the $\theta$ component. In physical terms, we conclude 
that for the large-$n$ families of fields, the flux rope 
in the field is wounded so tightly that it approaches the 
highly localized structure of a line current or a sheet current
outside of which $B_{\varphi}$ is negligible. In other words,
the free energy is stored as $B_{\theta}$ in this exterior region
associated with the line or sheet current.

Applying inequality (53) to inequality (49), we get
\begin{equation}
H_R \le 4 \pi F_{\varphi}^{\frac{m}{m+1}}
     \left[\frac{16(m+1)(\frac{4}{3} E_{max}+ E_0)}
     {(m+2)^2~\gamma}\right]^{\frac{1}{2m+2}} ~,
\end{equation}
\noindent
a tight bound on $H_R$.

To study this inequality, let us define
\begin{eqnarray}
C_1 & = & \left[\frac{16(m+1)}{(m+2)^2}\right]
^{\frac{1}{2m+2}}  ~, \\
C_2 & = & \left[\frac{16(m+1)(\frac{4}{3} E_{max}+ E_0)}
     {(m+2)^2~\gamma}\right]^{\frac{1}{2m+2}}
     =  C_1 \left(\frac{\frac{4}{3}E_{max}+E_0}{\gamma}
     \right)^{\frac{1}{2m+2}} ~.
\end{eqnarray}
\noindent Inequality (54) now can be written as
\begin{equation}
H_R \le 4 \pi F_{\varphi}^{\frac{m}{m+1}} C_2 ~.
\end{equation}

The top panel of Figure 5 shows the $C_2$ values, plotted
against the azimuthal fluxes, for $n=5, 7, 9$ families of fields.
We see that as $n$ increases from $n=5$ to $n=9$, the modulation 
of $C_2$ values decreases, with the $C_2$ values approaching $1$.
This indicates that as $n \rightarrow \infty$, $C_2$ values tend to 1.
This can be understood by the following analysis.
$C_2$ is a multiplication of $C_1$ and
$\left(\frac{\frac{4}{3}E_{max}+E_0}{\gamma}\right)^{\frac{1}{2m+2}}$.
Since $E_{max}$, $E_0$ and $\gamma$ are all bounded, the term 
$\left(\frac{\frac{4}{3}E_{max}+E_0}{\gamma}\right)^{\frac{1}{2m+2}}$
goes to 1 as $m \rightarrow \infty$ except for $\gamma=0$.
When $\gamma=0$, $H_R=0$. This is the trivial potential field
and is not an interesting field for our consideration
of helicity upper bound.
Owing to its mathematical form in equation (55),
$C_1$ approaches 1 as $m \rightarrow \infty$.
This is indicated by the plot in the bottom panel of
Figure 5, where $C_1$ values are plotted against $m$
and are approaching to 1 as $m$ increases.
So, on the limit of $n \rightarrow \infty$ or, equivalently,
$m \rightarrow \infty$,
\begin{eqnarray}
C_1 & \rightarrow & 1 \nonumber ~,\\
C_2 & \rightarrow & 1 \nonumber ~,\\
F_{\varphi}^{\frac{m}{m+1}}
     & \rightarrow & F_{\varphi} ~,
\end{eqnarray}
\noindent and hence we have
\begin{equation}
H_R \le 4 \pi F_{\varphi} ~.
\end{equation}

Now we see that in the limit of $n \rightarrow \infty$, the bound
on $H_R$ reduces to a problem of finding an upper bound on
$F_{\varphi}$. The force-free solutions of Flyer et al. (2004) 
suggest that such an upper bound exists (see also Figure 2 
in this paper). Our tighter inequality (59) has thus provided 
a more stringent bound on $H_R$ that had eluded from 
the lose inequality (35).

Figure 6 shows the variations of $H_R/(4 \pi F_{\varphi})$
against $\gamma$ for $n=5, 7$ and $9$ families.
These plots show that the relationship
$H_R \le 4 \pi F_{\varphi}$ 
we derived for $n \rightarrow \infty$ is also valid for
$n=5, 7, 9$ families. So this relationship may be a general
relationship that is valid for all $n$ values.
If this is true, it tells us that the upper bound on
the total magnetic helicity is a less stringent upper
bound than that on the total azimuthal flux.


\section{Conclusions and Discussions}

In this paper we proposed a hydromagnetic conjecture that
there is an upper bound on the total magnetic helicity
that a force-free field in an unbounded domain can contain.
Observations have shown that accumulations of helicity with 
a preferred sign takes places in the two solar hemispheres.  
If our conjecture is valid, the relentless accumulation of 
magnetic helicity in the corona will lead to a CME-type 
eruption as a natural and unavoidable product of coronal 
evolution. This conjecture deserves further investigation 
to find a mathematical rigorous proof, or disproof.  
For the present, evidence for and implications of this 
conjecture can be found in the several families of 
power-law axisymmetric force-free fields governed 
by equation (20).

We derived two rigorous inequalities (S1 and S2 in Table 1)
by applying the Cauchy-Schwartz and Holder inequalities, respectively.
The former (S1 in Table 1 or Equation (35) in the text) puts
an absolute upper bound on the magnitude of total magnetic helicity
for each fixed index $n$. This inequality is simple and rigorous
but it is not tight for our purpose, with
the bounds increasing monotonically with $n$.
The second inequality (S2 in Table 1) relates total
magnetic helicity, azimuthal flux and magnetic energy
together in an interesting way, which presents a more stringent
bound than S1 does.

Inequality S2 is further reduced to the simple
form of Inequality S3 in Table 1 when $n$ goes to infinity.
We have not proved that the simple form of Inequality S3 is valid
for all $n$ although the numerical data of Figure 6 suggest it is.
The validation of Inequality S3 for large $n$ is sufficient for
our purpose, which is to remove our main concern arising from
Inequality S1 of whether the total helicity is still bounded
when $n$ becomes very large. Inequality S3 together with the numerical 
results from Flyer et al. (2004) suggest that such an upper
bound on the total magnetic helicity exists even when $n$
increases without bound.

\begin{deluxetable}{cccccccc}
\tablecolumns{8}
\tablewidth{0pc}
\tablecaption{Significant equation and inequalities 
   derived in the paper}
\tablehead{ Number &  \multicolumn{5}{c}{Equation or Inequality}
  & \multicolumn{2}{c}{No. in the paper}}
\startdata
S0 & \multicolumn{5}{c}{$H_R = 4 \pi \int_{r>1} A
\times B_{\varphi} r dr d \theta$}
& \multicolumn{2}{c}{(Equ. 17)} \\
\\
S1 & \multicolumn{5}{c}{$-8 \pi \left[ {n+3 \over 2} E_0 +
\sqrt{{2 \gamma \over n+1}} c_0 \right]
\le H_R \le 8 \pi \left[{n+3 \over 2} E_0
- \sqrt{{2 \gamma \over n+1}} c_0 \right]$}
& \multicolumn{2}{c}{(Equ. 35)} \\
\\
S2 & \multicolumn{5}{c}{$H_R \le 4 \pi F_{\varphi}^{\frac{m}{m+1}}
     \left[\frac{16(m+1)(\frac{4}{3} E_{max}+ E_0)}
     {(m+2)^2~\gamma}\right]^{\frac{1}{2m+2}}$}
& \multicolumn{2}{c}{(Equ. 54)} \\
\\
S3 & \multicolumn{5}{c}{$H_R \le 4 \pi F_{\varphi}$}
& \multicolumn{2}{c}{(Equ. 59)} \\
\enddata
\end{deluxetable}

An important result of our study is the simple formula (S0 in Table 1)
for the relative helicity of an axisymmetric field.   This
formula is useful to future studies of axisymmetric fields
applicable, independent of whether the field is
force-free or in force balance. It has a simple 
physical interpretation: The total relative magnetic
helicity in axisymmetric fields is just a simple convolution
of the local wrapping of the azimuthal flux element
$B_{\varphi} r dr d \theta$ with the poloidal flux $A$.

We note that in Hu et al. (1997) the right hand side of
Equation S0 has been defined to be a form of helicity and
the authors have showed that this form of helicity is
conserved under the ideal hydromagnetic induction equation.
Their derivation was carried out completely independent of
the concept of relative helicity introduced by Berger
\& Field (1984). However, our derivation of Equation S0
here, from a direct application of the concept of the Berger
relative helicity, shows that the helicity defined
by Hu et al. (1997) is actually the same physical quantity
as the Berger relative helicity for an axisymmetric field.

What is the relationship between magnetic helicity 
accumulation and free magnetic energy storage?
Which one plays a more fundamental role in producing CMEs?
It seems clear that the greater the total helicity the greater
the magnetic energy would be, as suggested by the inequality
(2).  Although this inequality is rigorous only for a
finite domain, a similar conclusion is expected generally 
in the unbounded domain that higher twist implies greater
magnetic energy.  The fact that the magnetic energy of a 
force-free field in the unbounded domain has an upper bound 
suggests that our conjecture on a similar kind of bound
on the total helicity is probably physically sensible.
It is interesting to observe from our Figure 2 that
the maximum storages of magnetic helicity and magnetic energy
do not occur at the same values of $\gamma$.
We suggest that magnetic helicity as a conserved quantity
is for sure to be accumulated in the corona whereas the corona
may release its free energy by magnetic reconnection without
leading to a CME expulsion. So for fields with large energy 
storage but moderate helicity storage, the field may release 
energy when a suitable trigger acts on it. But when magnetic 
helicity has accumulated to cross over the upper bound, 
we proposed, a CME expulsion becomes unavoidable.

In this sense, exceeding
the allowable level of helicity in a force-free field is
only a sufficient condition for an eruption. 
It is an extreme condition where a field has accumulated
enough helicity that an eruption becomes unavoidable. A CME expulsion
may still occur even before this helicity limit is reached,
as long as the separate necessary condition, that of having enough
magnetic energy for an eruption, is met. For example, mass-loading
is an efficient way to store enough
free magnetic energy to drive CMEs, without having to cross
the threshold on excessive magnetic helicity accumulation.
In Zhang \& Low (2004) an equilibrium state is given where 
prominences are supported by the magnetic fields in
the so-called normal configuration. These solutions
show that significant magnetic energy can be stored
with only a relatively small amount of total azimuthal 
magnetic flux and magnetic helicity. An interesting
question for future investigation is then whether a coronal
field may accumulate helicity to the point of exceeding
the applicable upper bound without first having enough
free magnetic energy to erupt into a CME.


\acknowledgements

We thank Tom Bogdan and E. N. Parker for helpful comments.
We also thank the anonymous referee whose comments and
suggestions have improved the quality of this paper.
This work was partly supported by the One-Hundred-Talent
Program of Chinese Academy of Sciences, Chinese National 
Key Basic Research Science Foundation (G2000078404), 
Chinese National Science Foundation Grant 10373016, 
NASA Living with a Star Program, NSF/SHINE ATM-0203489
and NSF ATM-0548060 Programs. 
The National Center for Atmospheric Research
is sponsored by the National Science Foundation.


\clearpage
\begin{figure}
\epsscale{.75}
\plotone{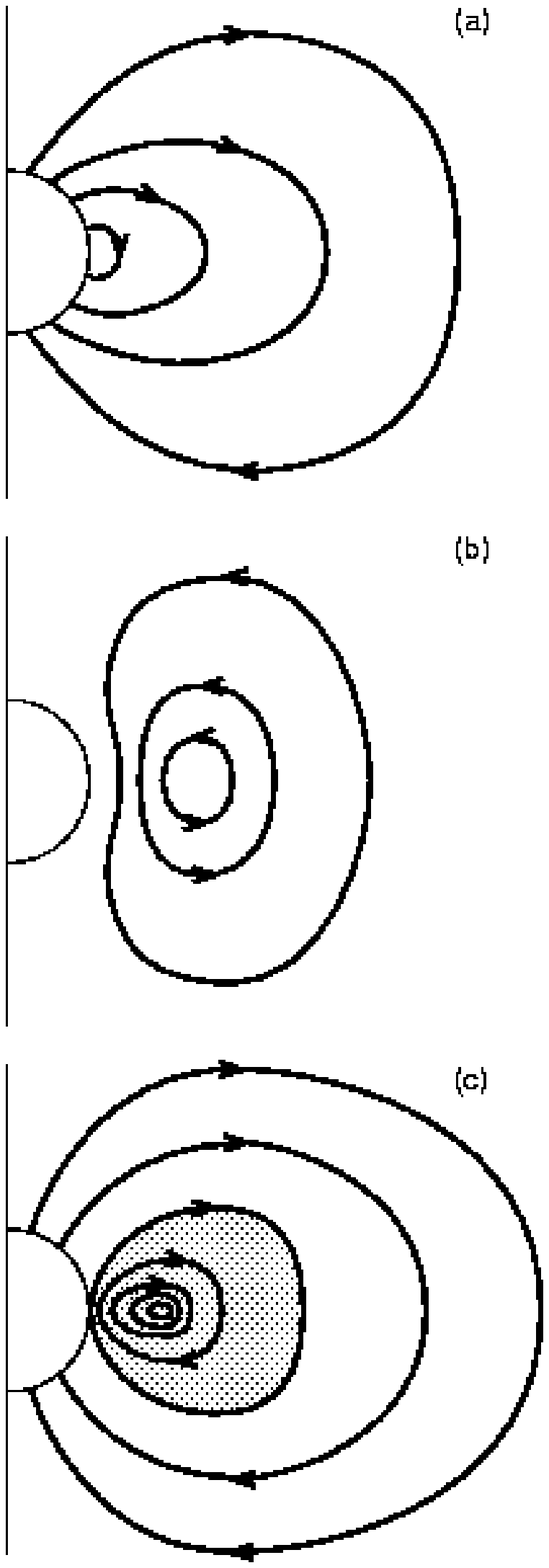}
\vspace{1.0cm}
\caption{\small{Shown is an axisymmetric magnetic field
with its lines of force seen projected on a $r-\theta$ plane,
taken from Low (2001). The field lines are directed out of
the shown plane if $B_{\varphi} \ne 0$.
The middle panel shows an azimuthal field with $B_r = 0$
on $r = 1$, which cannot find an equilibrium.
The upper panel shows a potential field with $B_r \ne 0$ on
$r = 1$, meeting the necessary condition for an equilibrium.
The bottom panel shows a field also meeting the necessary
condition but with a detached flux rope, represented by
the closed loops of poloidal projected lines.
This configuration can be regarded as a sum of the above
two configurations, of a field containing a certain amount
of magnetic helicity and finding an equilibrium.}}
\end{figure}

\begin{figure}
\epsscale{.7}
\plotone{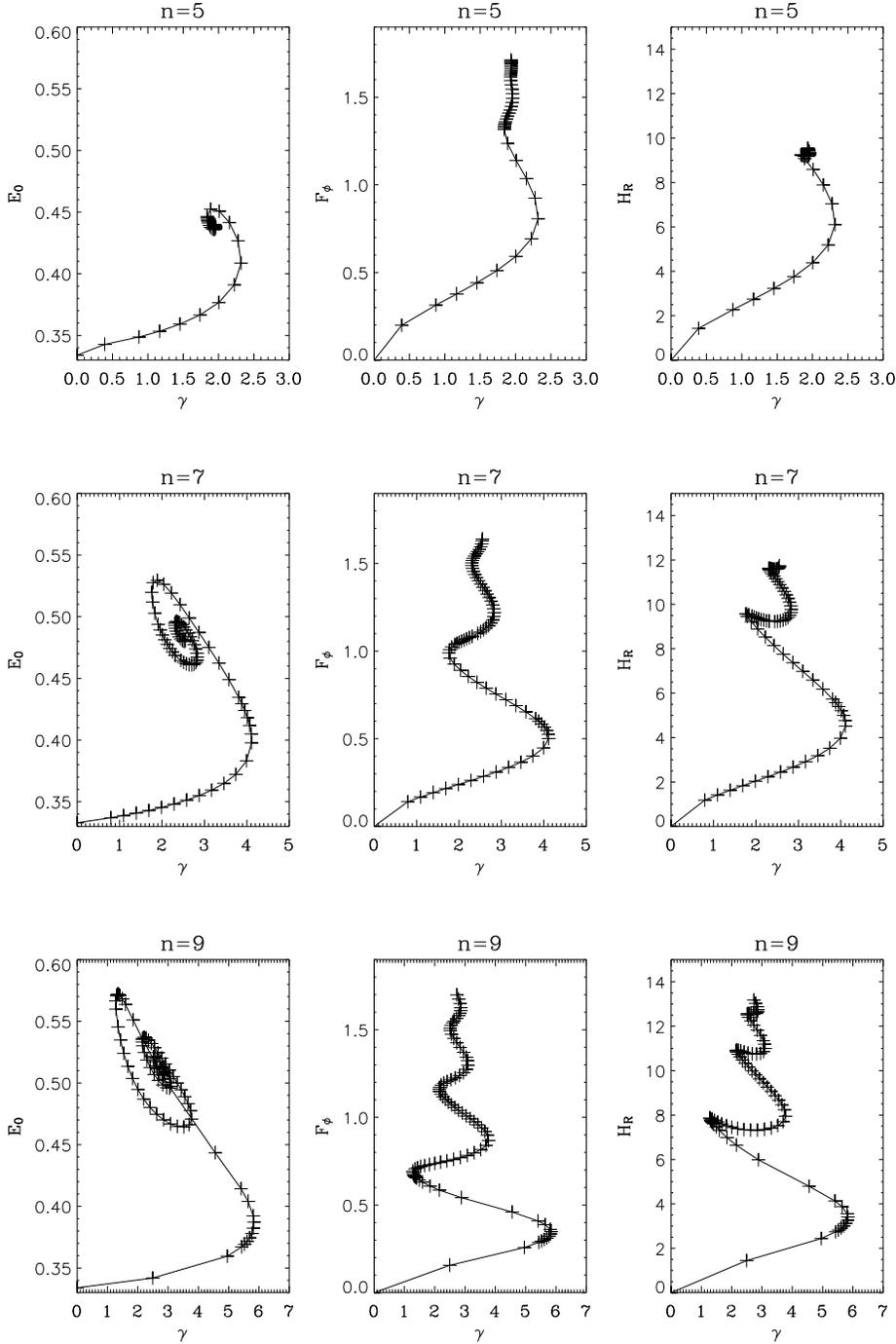}
\vspace{0.5cm}
\caption{\small{Variation of the total magnetic energy
($E_0$), azimuthal flux ($F_{\varphi}$) and
total magnetic helicity ($H_R$) vs. $\gamma$
along the solution curve for $n = 5$ (top panels), $n = 7$
(middle panels) and $n = 9$ (bottom panels).
Each point, denoted by a plus symbol in the figure, 
represents a solution to equation (20).
These curves of the solutions suggest that there may be
an upper bound on the total magnetic helicity in addition to
those on the total magnetic energy and azimuthal flux.
See Flyer et al. (2004) for details of the solutions
and the numerical solver.}}
\end{figure}

\begin{figure}
\epsscale{.7}
\plotone{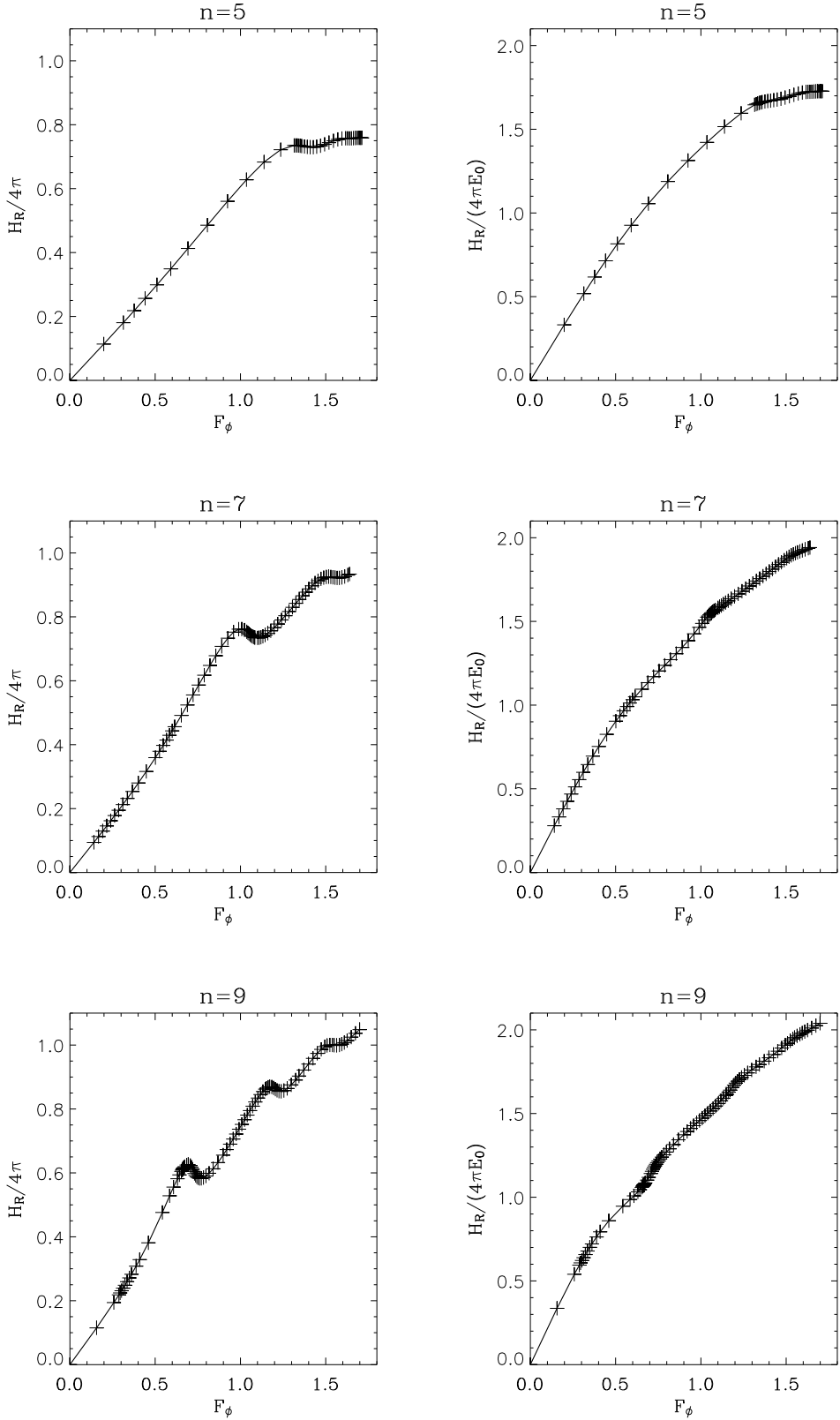}
\vspace{0.5cm}
\caption{\small{Diagrams of the total magnetic helicity $H_R$
(left panels) or of the total magnetic helicity divided
by the total magnetic energy $H_R/E_0$ (right panels) 
of each solution against the azimuthal flux $F_{\varphi}$.
Here $H_R$ has been normalized by a factor of $4\pi$.
See text for discussions.}}
\end{figure}

\begin{figure}
\plotone{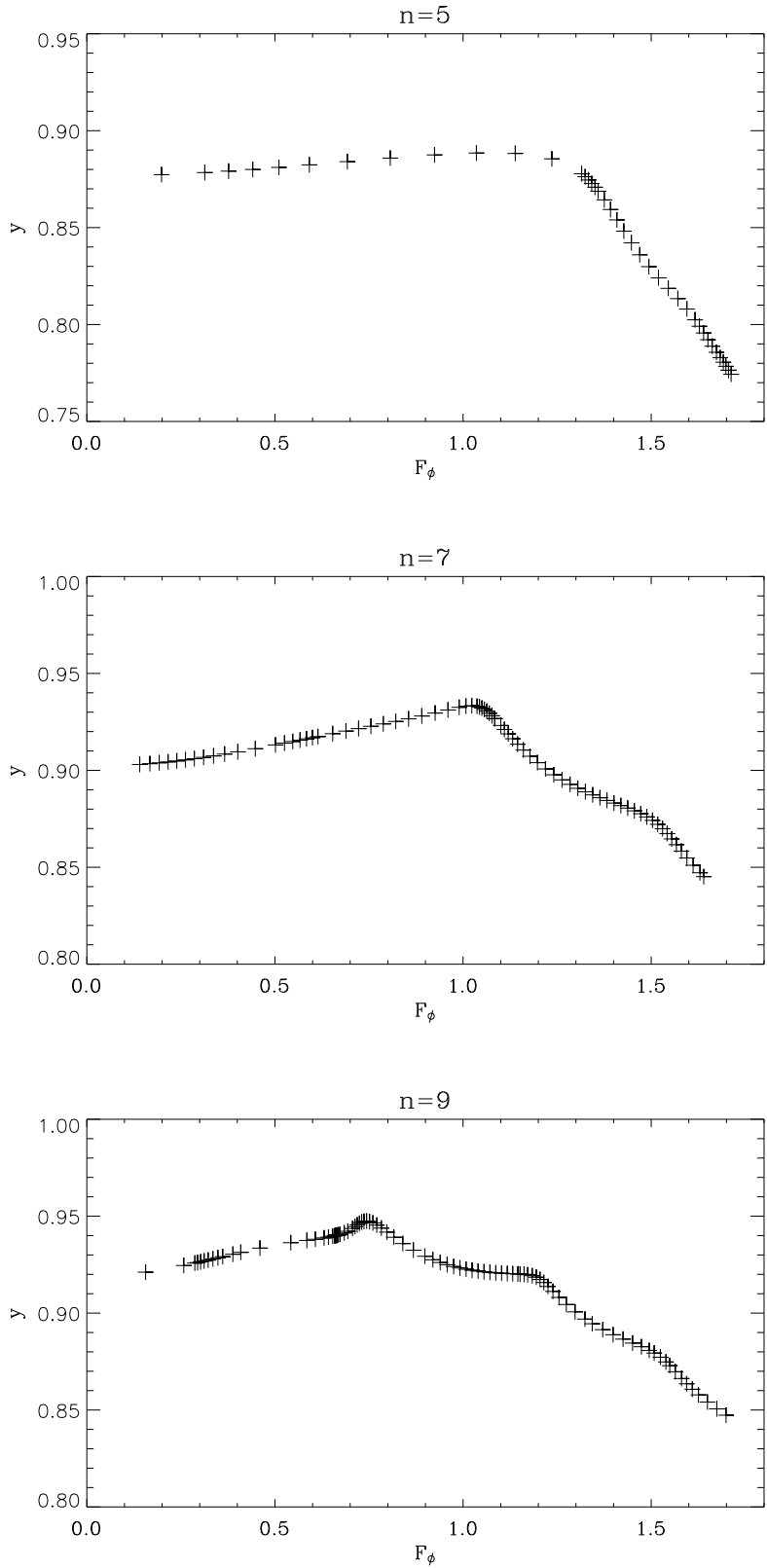}
\vspace{0.5cm}
\caption{\small{Plots of $y$ values against azimuthal fluxes
$F_{\varphi}$ for $n=5, 7, 9$ families of solutions. 
See text for definition and discussions.}}
\end{figure}

\begin{figure}
\epsscale{1}
\plotone{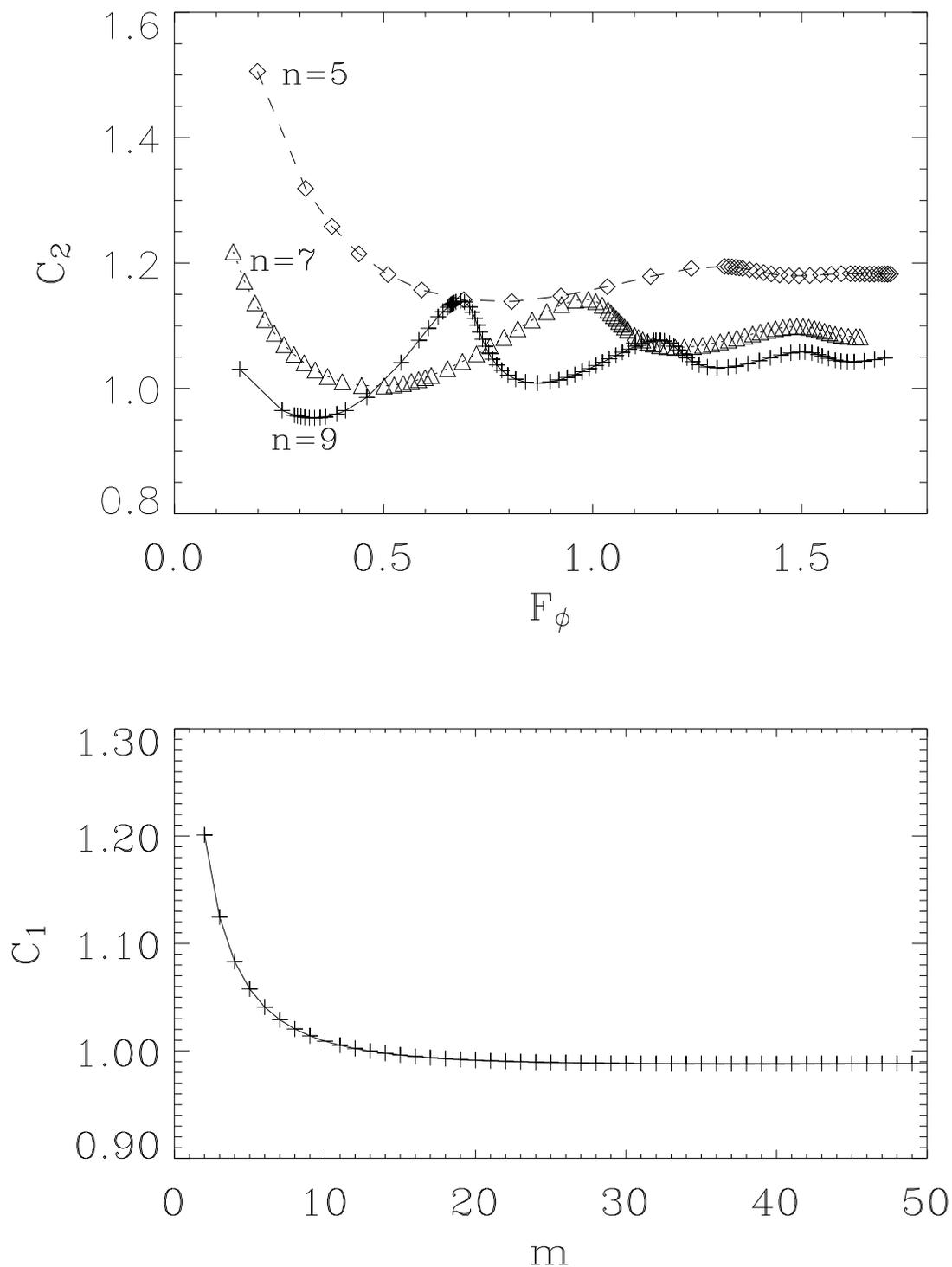}
\caption{\small{Top panel: Plots of $C_2$ values against 
azimuthal fluxes for $n=5, 7, 9$ families of solutions.
See text for definitions and discussions.
Bottom panel: Variation of $C_1$ values against $m$.
This plot shows that as $m$ goes to infinity
$C_1$ goes to 1.}}
\end{figure}

\begin{figure}
\plotone{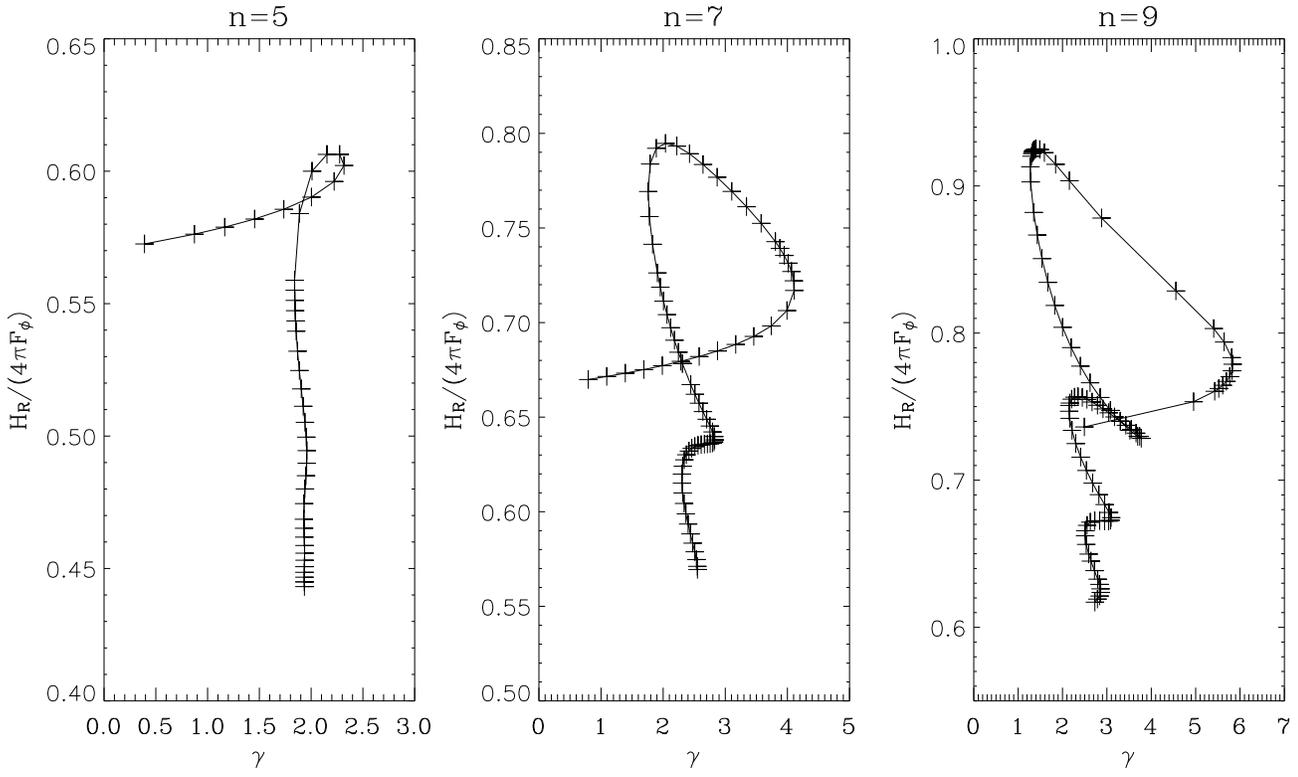}
\caption{\small{Plots of $H_R/(4\pi F_{\varphi})$
against $\gamma$ for $n=5, 7, 9$ families of solutions.
These plots show that the relationship
$H_R \le 4 \pi F_{\varphi}$ also valid for $n=5, 7, 9$
families, so it may be a general relationship
valid for all $n$ values.}}
\end{figure}


\begin{thebibliography}{}

\bibitem[Aly 1984]{aly84}
Aly, J. J. 1984, ApJ, 283, 349

\bibitem[Aly 1988]{aly88}
Aly, J. J. 1988, A\&A, 203, 183

\bibitem[Aly 1991]{aly91}
Aly, J. J. 1991, ApJ, 375, L61

\bibitem[Antichos et al. 1999]{antiochos99}
Antiochos, S. K., DeVore, C. R., \& Klimchuk, J. A. 1999,
ApJ, 510, 485

\bibitem[Bao \& Zhang 1998]{bao98}
Bao, S., \& Zhang, H. Q. 1998, ApJ, 496, L43

\bibitem[Berger 1984]{berger84}
Berger, M. A. 1984, Geophys. Astrophys. Fluid Dynamics, 
30, 79

\bibitem[Berger 1985] {berger85}
Berger, M. A. 1985, ApJ Suppl., 59, 433

\bibitem[Berger \& Field 1984]{berger84b}
Berger, M. A., \& Field, G. B. 1984, J. Fluid Mech., 147, 133

\bibitem[Burkepile et al. 2004] {burkepile04}
Burkepile, J. T., Hundhausen, A. J., Stanger, A. L., et al. 
2004, JGR, 109, 3103

\bibitem[Chandrasekhar 1961] {chandra}
Chandrasekhar, S. 1961, Hydrodynamic and Hydromagnetic
Stability (Oxford : Oxford Univ. Press)

\bibitem[Chen \& Shibata 2000]{chen00b}
Chen, P. F., \& Shibata, K. 2000, ApJ, 545, 524

\bibitem[Dere et al. 1999]{dere99}
Dere, K. P., Brueckner, G. E., Howard, R. A., 
Michels, D. J., \& Delaboudiniere, J. P. 1999,
ApJ, 516, 465

\bibitem[Flyer et al. 2004]{flyer04}
Flyer, N., Fornberg, B., Thomas, S., \& Low, B. C.
2004, ApJ, 606, 1210

\bibitem[Flyer et al. 2005]{flyer05}
Flyer, N., Fornberg, B., Thomas, S., \& Low, B. C.
2005, ApJ, 631, 1239

\bibitem[Fong et al. 2002]{fong02}
Fong, B., Low, B. C., \& Fan, Y. H. 2002, ApJ, 571, 987

\bibitem[Forbes \& Isenberg 1991]{forbes91}
Forbes, T. G., \& Isenberg, P. A. 1991, ApJ, 373, 294

\bibitem[Howard et al. 1976]{howard76}
Howard, R. A., Koomen, M. J., Michels, D. J., 
Tousey, R., Detwiler, C. R., et al. 1976,
NOAA World Data Center A for Solar-Terrestrial Phys.
Report, UAG-48A

\bibitem[Hu et al. 1997]{hu97}
Hu, Y. Q., Xia, L. D., Li, X., Wang, J. X., \& Ai, G. X. 
1997, Sol. Phys., 170, 283

\bibitem[Hu et al. 2003]{hu03}
Hu, Y. Q., Li, G. Q., \& Xing, X. Y. 2003,
J. Geophys. Res., 108, 1072

\bibitem[Hudson \& Cliver 2001]{hudson01}
Hudson, H. S., \& Cliver, E. W. 2001,
J. Geophys. Res., 106, 25199

\bibitem[Hundhausen 1999] {hundausen99}
Hundhausen, A. J., 1999, in The Many Faces of the Sun,
ed. K. Strong, J. Saba, B. Haisch \& J. Schmelz
(New York : Springer), 143

\bibitem[Isenberg et al. 1993]{Isenberg93}
Isenberg, P. A., Forbes, T. G., \& Demoulin, P. 1993,
ApJ, 417, 368

\bibitem[Kurokawa 1987]{kurokawa87}
Kurokawa, H. 1987, Sol. Phys., 113, 259

\bibitem[Leka et al. 1996]{leka96}
Leka, K. D., Canfield, R. C., McClymont, A. N., 
\& van Driel-Gesztelyi, L. 1996, ApJ, 462, 547

\bibitem[Lite et al. 1995]{lite95}
Lites, B. W., Low, B. C., Martinez-Pillet, V., Seagrave, P.,
Skumanich, A., et al. 1995, ApJ, 446, 877

\bibitem[Low 2001] {low01}
Low, B. C. 2001, J. Geophys. Res., 106, 25141

\bibitem[Low 2003] {low03}
Low, B. C., Fong, B., \& Fan, Y. H. 2003, ApJ, 594, 1060

\bibitem[Low \& Smith 1993]{low93}
Low, B. C., \& Smith, D. F. 1993, ApJ, 410, 412

\bibitem[MacQueen et al. 1974]{macqueen74}
MacQueen, R. M., Eddy, J. A., Gosling, J. T., Hilder, E., 
Munro, R. H., et al. 1974, ApJ, 187, L85

\bibitem[Moffatt 1985]{moffatt85}
Moffatt, H. K. 1985, J. Fluid Mech., 159, 359

\bibitem[Parker 1994]{parker94}
Parker, E. N. 1994, Spontaneous Current Sheets in 
Magnetic Fields, New York : Oxford U. Press

\bibitem[Pevtsov et al. 1995]{pevtsov95}
Pevtsov, A. A., Canfield, R. C., \& Metcalf, T. R. 1995,
ApJ, 440, L109

\bibitem[Pevtsov et al. 2003]{pevtsov03}
Pevtsov, A. A., Maleev, V. M., \& Longcope, D. W. 2003,
ApJ, 593, 1217

\bibitem[Rust 1994]{rust94}
Rust, D. M. 1994, Geophys. Res. Lett., 21, 241

\bibitem[Webb \& Hundhausen 1987]{webb87}
Webb, D. F., \& Hundhausen, A. J. 1987, Sol. Phys., 108, 383

\bibitem[Wolfson 2003] {wolfson03}
Wolfson, R. 2003, ApJ, 593, 1208

\bibitem[Wolfson \& Dlamini 1997] {wolfson97}
Wolfson, R. \& Dlamini, B. 1997, ApJ, 483, 961

\bibitem[St. Cyr et al. 2000] {st00}
St. Cyr, O. C., et al. 2000, J. Geophys. Res., 105, 18169

\bibitem[Sturrock 1991]{Sturruck91}
Sturrock, P. A. 1991, ApJ, 380, 655

\bibitem[Woltjer 1958]{woltjer58}
Woltjer, L. 1958, Proc. US Natl. Acad. Sci., 44, 489

\bibitem[Zhang 2001]{zhangh01}
Zhang, H. Q. 2001, MNRAS, 326, 57

\bibitem[Zhang and Low 2001] {zhang01}
Zhang, M., \& Low, B. C. 2001, ApJ, 561, 406

\bibitem[Zhang and Low 2003] {zhang03}
Zhang, M., \& Low, B. C. 2003, ApJ, 584, 479

\bibitem[Zhang and Low 2004] {zhang04}
Zhang, M., \& Low, B. C. 2004, ApJ, 600, 1043

\bibitem[Zhang and Low 2005] {zhang05}
Zhang, M., \& Low, B. C. 2005, ARAA, 43, 103

\end{thebibliography}
\end{document}